# On the Difficulties of Incentivizing Online Privacy through Transparency: A Qualitative Survey of the German Health Insurance Market

Max Maass[1], Nicolas Walter[1], Dominik Herrmann[2], and Matthias Hollick[1]

[1] TU Darmstadt, Secure Mobile Networking Lab, Darmstadt, Germany
{mmaass,nwalter,mhollick}@seemoo.tu-darmstadt.de
[2] University of Bamberg, Privacy and Security in Information Systems, Bamberg, Germany
dominik.herrmann@uni-bamberg.de

**Abstract.** Today, online privacy is the domain of regulatory measures and privacy-enhancing technologies. Transparency in the form of external and public assessments has been proposed for improving privacy and security because it exposes otherwise hidden deficiencies. Previous work has studied privacy attitudes and behavior of consumers. However, little is known on how organizations react to measures that employ public "naming and shaming" as an incentive for improvement. We performed the first study on this aspect by conducting a qualitative survey with 152 German health insurers. We scanned their websites with PrivacyScore.org to generate a public ranking and confronted the insurers with the results. We obtained a response rate of 27%. Responses ranged from positive feedback to legal threats. Only 12% of the sites – mostly non-responders – improved during our study. Our results show that insurers struggle due to unawareness, reluctance, and incapability, and demonstrate the general difficulties of transparency-based approaches.

**Keywords:** privacy, web security, transparency, quasi-experiment, e-mail survey

## 1 Introduction

Privacy plays an increasingly important role for consumers, regulators, and companies, especially on the internet. A growing number of companies provide services in the areas of online advertising, web analytics, and user profiling. Users and regulators have responded by deploying tracking blockers and passing stronger privacy laws, e.g., the European General Data Protection Regulation (GDPR).

So far, most research has focused on user perceptions of information privacy [1–4] and models for firms' data sharing gathering [5–7] and sharing behavior [8]. However, there is a lack of research on how to incentivize companies to reduce their use of tracking services, which are disliked by many users [9]. The study presented in this paper investigates the role of *transparency* as an incentive mechanism. Our scope is not limited to *privacy* (represented by online tracking services); we also consider the connected area of *security* (e.g., transport encryption of web and mail traffic), as they affect user privacy against malicious actors like criminals and intelligence agencies.

Existing website scanners like the Qualys SSL Test (ssllabs.com) and Webbkoll [10] allow to assess security and privacy features of single sites. However, these scanners communicate the result only to the particular user who commissioned a scan. Therefore, there is little incentive for site operators to improve their rating. In contrast, we consider a scenario in which the scan results for multiple competing organizations are published openly on the internet in a ranking. This combination of publicity and comparability creates more transparency for consumers, which may increase the pressure on site operators to improve. Therefore, we pose the following research questions:

- **RQ1**: How do website operators react when they are notified that their website has been rated in terms of privacy and security aspects and the results are publicly available online?
- **RQ2**: Does telling them about being in a *public ranking* change their reaction?

We seek to answer these questions by contacting 152 German health insurance providers in a qualitative study, confronting them with and asking their comment on the current state of information privacy on their websites as determined by the PrivacyScore platform [11]. PrivacyScore.org is a public web service that automatically analyzes websites for security and privacy issues. Since its inception in June 2017, it has performed over 1 million scans and is being used by activists, data protection officers, and the general public. To the best of our knowledge, our study is the first to investigate how the competitive nature of public privacy and security rankings affects website operators, using the ranking functionality offered by PrivacyScore.

## 2 Related Work

The majority of IS privacy research focuses on individual privacy [6]. One of the central questions in this area is how privacy concerns affect the way individuals behave and their willingness to disclose personal information [1]. A frequently discussed aspect in this context refers to the decisions that individuals make regarding privacy and a trade-off between risk and benefit [3], which is informed by many factors, including company culture, legal environment, and the industry of the information-gathering company [4].

In contrast, there is less research on the question of how privacy issues are perceived from the **organization perspective**, i.e., by companies who work with personal data of their consumers. This field concerns itself with the interests and attitudes and motivations that companies have towards privacy [1, 5–8]. Greenaway and Chan argue that companies can behave in a reactive or proactive manner in regard to privacy and thus influence the perception of customers, and that their behavior can be explained using two models [5]. The *Institutional Approach* (IA) considers firms behavior as a search for legitimacy in the face of external pressures, and distinguishes an *acquiescent* (compliance with law, imitation of peers) and a *proactive approach* (exceeding minimum requirements and using clear communication to achieve leadership without falling out of line). The *Resource-based View* (RbV) posits that firms seek a sustainable competitive advantage using either an *information-* (through superior data analysis) or a *customer focus* (through superior customer trust).

To the best of our knowledge, the role of **competition** in promoting privacy has not received significant attention so far, although individual companies like Apple[1] or DuckDuckGo[2] have started marketing themselves as privacy champions. Like the RbV, Ohlhausen and Okuliar view privacy as a factor in competition [12]. Kerber reverses the argument and posits that a lack of competition may be partially responsible for the lack of privacy offerings from online companies [13].

Another reason may be that the privacy behavior of companies is often invisible from the outside. Thus, consumers find it hard to check whether companies are living up to their claims, for instance by following accepted best practices that ensure the privacy and security of their customers. **Transparency** can make these invisible practices of companies visible, enabling customers to easily verify claims and compare the practices of competing companies. The IS transparency literature has so far been focused on transparency strategies for companies [14], but has not yet considered transparency provided by external actors in the field of information privacy. Such external transparency has been shown to improve outcomes in areas such as Corporate Social Responsibility [15] and corporate ethics [16], among others, making it a promising avenue for more research.

## 3    Technical Background

We use the platform PrivacyScore.org [11] to investigate how operators of websites react when they are confronted with the fact that a third party is creating transparency about the privacy and security practices exhibited by their websites.

PrivacyScore.org[3] is an automated website scanner that allows anyone to investigate websites regarding privacy and security issues. Compared to other website scanners like Qualys SSL Scan and Webbkoll, the scans conducted by PrivacyScore are more comprehensive, including both security and privacy aspects. Currently, PrivacyScore performs more than 60 checks in four groups: *Tracking and Privacy* checks if a website includes third-party trackers. *Website Encryption* checks if a web server offers state-of-the-art secure connections, whether it redirects users to the secure version of the site, and whether it sets the HTTP Strict Transport Security (HSTS)[4] header, which prevents certain classes of attacks. *Mail Encryption* checks whether the mail servers of a website support state-of-the-art encryption, and *Web Security* checks if a website leaks internal information [17] and whether it protects users against client-side attacks.

Detailed results of each scan are published on the website and retained in a database to facilitate longitudinal studies. While it is possible to scan individual websites with PrivacyScore, it is primarily intended to be used to scan lists of related websites (e.g., all pharmacies with an online shop in a specific country). Scanning a list generates a public ranking with all sites that belong to a list. Rankings put the individual scan results into context and allow consumers, data protection authorities, and site operators to assess how a particular site compares with its competitors. At the time of writing PrivacyScore hosts over 200 scan lists and has conducted more than 1 million scans.

---

[1] https://www.cnbc.com/2018/10/25/in-privacy-fight-apple-has-more-to-lose-than-facebook-or-google.html (accessed 2018-11-16)
[2] https://www.theverge.com/2017/1/25/14381138/duckduckduckgo-privacy-anonymous-searching-10-billion (accessed 2018-11-16)
[3] The first and third authors of this paper are members of the PrivacyScore team.
[4] https://tools.ietf.org/html/rfc6797 (accessed 2018-11-19)

A noteworthy limitation of PrivacyScore is that it conducts scans automatically without human oversight. As a result, the assessment is limited to aspects that can be reliably verified programmatically from the outside by visiting the website with an instrumented web browser. In particular, PrivacyScore cannot verify whether statements in the privacy policy are adequate and whether the site actually adheres to them. Moreover, PrivacyScore conducts only benign checks. For instance, it does not actively attempt to find and exploit vulnerabilities in the website under test.

Finally, PrivacyScore attempts to find a fair balance between satisfying the desire for transparency of consumers and operational interests of site operators. To this end, all scans are conducted instantaneously, i.e., without asking operators for permission. Website operators can have their website excluded from future scans. For reasons of transparency, this fact is published next to the scan results. Moreover, the last successful scan result before the complaint is shown for the respective site.

## 4  Study Design

As proposed by Greenaway and Chan [5], we use a mixed method [18] approach by combining a classical open question survey with a quasi-experimental setup based on the PrivacyScore platform. We follow a qualitative research process consisting of planning, data gathering, preparation of analysis, and analysis and summarization [19].

### 4.1  Gathering Websites and Scan Results

Our focus is on the health insurance sector within Germany, as visitors of health insurance websites may disclose sensitive information (e.g., by looking up information about specific illnesses and treatments). We obtained the homepage addresses (domain names) of all public and private health insurers in Germany from Wikipedia,[5] uploaded the URLs to PrivacyScore and added them to the newly created scan list "Deutsche Krankenkassen und Krankenversicherungen" (privacyscore.org/list/15). In December 2017 (T1) we saved a local copy of the scan result for each website as well as its rank within the aforementioned list. We performed two additional scans at later points in time (February and August 2018; T2 and T3) to determine whether the insurers had changed security and privacy properties of their sites.

---

[5] https://de.wikipedia.org/w/?title=Liste_deutscher_Krankenkassen&oldid=180861868 and https://de.wikipedia.org/w/?title=Liste_deutscher_privater_Krankenversicherer&oldid=180752925 (accessed 2018-09-11)

### 4.2 Group Allocation

Seeking to answer RQ2 we use a quasi-experimental setup, which is common in organizational research [20]. We split the insurers into two groups A and B, introducing the variable *competition*. We prepared two kinds of solicitation emails. Insurers in Group A were only informed about the overall scan results for their site in terms of the four areas that are published on the PrivacyScore website: "Tracking and Privacy", "Website Encryption", "Mail Encryption", and "Web Security". Insurers in Group B were additionally confronted with the rank of their site in our "Deutsche Krankenkassen und Krankenversicherungen" scan list. All solicitation emails started out with a brief description of PrivacyScore and ended with a request to participate in our study by answering the following two open questions: *What do you think about this kind of an assessment from a company's point of view? Would you consider making changes to your website in order to improve its privacy properties?*

The group allocation was not randomized but based on the ranking of the websites on PrivacyScore. This was done to ensure that both groups are homogenous in terms of the distribution of the ranks as given by the scan list. To this end, we assigned the numbers from 1 to 152 to the insurers in the order of the PrivacyScore ranking. The companies with odd numbers were put into Group A, those with even numbers into Group B. This procedure was guaranteed to result in two groups of the same size and allowed us to handle cases of ties in a deterministic fashion. Note that our allocation resulted in an uneven distribution in terms of the *type of insurer*: Group A contained 61 public and 15 private insurers, whereas Group B contained 51 and 25, respectively. This heterogeneity will be taken into account during the analysis.

### 4.3 Sending of Solicitation Mails

We obtained an email address for every insurer by visiting each website and manually searching for a suitable contact address in the privacy policy, the contact section, and the imprint of the site. If available, we preferred data protection-specific addresses such as "datenschutz@" to general-purpose addresses like "service@" or "info@".

On 8 December 2017, we sent out solicitation emails to all 152 insurers, using a university mail account not affiliated with the PrivacyScore project. As most insurers had not replied with a definitive response after four weeks, we sent out reminders to 135 insurers on 11 January 2018. The majority of responses was received in January and February 2018. After we had received all responses (March 2018) we iterated over them multiple times for open coding. As two of the authors are part of the PrivacyScore team, we were also in the position to receive any complaints directed towards the platform (which was not communicated to the recipients).

## 5 Results

In this section, we report the results of the initial scan of the websites and how site operators responded to our solicitation mails. Afterwards we discuss the three primary response types, i.e., positive responses, complaints, and other responses, in more detail.

### 5.1 Initial State of Websites

At the beginning of the study (T1), 26% (39) of the websites did not use any third-party tracking. 67% (102) of the websites used a well-configured Transport Layer Security (TLS) setup (defined as automatically forwarding users to the encrypted version of the website, offering TLS 1.2 and not offering the outdated and insecure protocol versions SSLv2 and SSLv3), and only 26% (39) of the websites set a HTTP Strict Transport Security (HSTS) header. Mail server TLS scans exhibited a failure rate of 6.5% (15), likely due to spam protection techniques like tarpitting. Still, 78% (119) websites had a well-configured mail server TLS setup (with TLS 1.2 and without SSLv2 and SSLv3).

### 5.2 Contacts and Respondents

In total, we received responses from 41 insurers (response rate: 27%). We found that sending a reminder message greatly increased the response rate, with more than half of the responses (23 of 41) reaching us only after the reminder. We had contacted 97 data protection (DP) contacts (64% of 152 insurers; A=45, B=52) and 55 general-purpose contacts (A=31, B=24). The overall response rates of Group A and B are similar (A=26%, B=28%), with data protection contacts showing a slightly higher rate (A=29%, B=31%) than general-purpose contacts (A=23%, B=21%).

We also observed many instances of messages being forwarded inside the companies, with different departments responding to our inquiries. An overview is given in Table 1. In Group A, our messages were frequently answered by marketing and data protection or IT teams (40% and 35%, respectively). Members of the board of directors (20%) and other departments (5%) sent the remaining 25% of responses.

In Group B, we observed a markedly different behavior. The majority of the responses were written by data protection and IT specialists (71%), while the remainder was evenly split between marketing, board members, and others. This deviation can only partially be explained by the larger number of DP contacts in Group B (A=45, B=52): The fraction of data protection and IT responses is higher in Group B for both messages sent directly to a DP contact (A=35%, B=75%) and for messages sent to a general-purpose contact (A=14%, B=60%).

### 5.3 Types of Responses

Overall, the responses fall into one of the following three categories: Firstly, there are *positive responses*, e.g., statements of gratefulness, expressions of interest in the study, and very detailed discussions of the scan results. Secondly, there are *complaints* criticizing the nature of the unsolicited scans and the publication of the scan results. Thirdly, there are *other responses*, ranging from acknowledgements of the receipt of our message to explicit expressions of indifference. The heat map in Fig. 1 provides a summary of the results by breaking down the responses according to their response type. It also indicates the relationship between response type and changes to the website, which we will analyze in Sect. 5.4 in more detail. All direct quotes in this paper are our translations of German replies.

**Table 1.** Recipients and Respondents

| Group A | | | Group B | | |
|---|---|---|---|---|---|
| *Sent to* | *Response From* | *Count* | *Sent to* | *Response From* | *Count* |
| General | DPO / IT | 1 | General | DPO / IT | 4 |
| | Marketing | 3 | | Marketing | 1 |
| | Board Member | 1 | | | |
| | Other | 2 | | | |
| DPO | DPO / IT | 6 | DPO | DPO / IT | 12 |
| | Marketing | 5 | | Marketing | 1 |
| | Board Member | 1 | | Board Member | 2 |
| | Other | 1 | | Other | 1 |

**Positive Responses.** In total, we received 16 positive responses (A=9, B=7). In Group A, most responses came from the marketing (5) and data protection / IT departments (3), with a single response from a board member. In Group B, the responses came almost exclusively from data protection or IT departments (6), with only a single response from a marketing department. The responses varied widely in length and level of detail, ranging from single-sentence responses thanking us for the information and noting that the IT department would be investigating the report further, to differentiated technical and economic analyses of the trade-offs between user privacy and economic success for their company. For example, $B_{58}$ explained that while they used tracking services on their websites, they were disabled in sensitive areas, and noted that "*tracking is above all about the care with which data is handled. The mere collection of data does not necessarily lead to better business success.*" $A_{33}$ noted that while they did not forward users to the secure version of their website by default, all sensitive data (login information, payment details) were transmitted over a secure connection.

Two respondents claimed that they would be making changes based on the results of the scan: $A_{67}$ gave a detailed response, referring to many individual test results in detail, and demonstrated that they had already reacted by implementing some changes, like enabling the Referrer-Policy[6] HTTP header. They also noted that some parameters related to the mail server were outside of their control, as they were managed by a third-party appliance. $A_{63}$ responded that internal tests had confirmed the findings of PrivacyScore and mentioned nonspecific changes that were made to the website as a result. They also noted that "*the results helped us protect the privacy of the users of our website [and provided] valuable support for the implementation of changes based on an easier analysis and identification of weak points.*" In contrast, $A_{35}$ noted the beta-status of PrivacyScore, but claimed that they would be performing their own checks to confirm the results, and act upon them, if necessary.

---

[6] https://www.w3.org/TR/referrer-policy/ (accessed 2018-11-19)

| Response Type | | Insurer Type | | Group | | Website improved? | |
|---|---|---|---|---|---|---|---|
| | Total | Public | Private | Group A | Group B | yes | no |
| Positive | 11 % | 7 % | 3 % | 6 % | 5 % | 1 % | 10 % |
| Neutral | 11 % | 9 % | 2 % | 5 % | 5 % | 0 % | 11 % |
| Complaint | 6 % | 6 % | 0 % | 2 % | 4 % | 0 % | 1 %* |
| None | 73 % | 52 % | 21 % | 37 % | 36 % | 11 % | 66 % |
| Sum | 100 %** | 74 % | 26 % | 50 % | 50 % | 12 % | 88 % |

*No scan results for 8 insurers that opted to be excluded from further scans. **Deviation due to rounding error.

**Figure 1.** Breakdown of responses according to response type

Finally, $A_{91}$ provided a highly detailed response, going into some detail on the trade-offs between user privacy and economic success in online tracking. They noted that online tracking was required to evaluate the effectiveness of their online affiliate marketing campaigns. An additional concern was the analysis of the needs of potential customers, noting that "*we determine the needs of new and existing customers through market research studies. However, [...] their results are only of limited significance. This can be seen in the fact that statements determined by market research and actual user behavior partially contradict each other.*" Another concern was personalization, with the respondent noting that "*both market research and measurements show that personalized content [...] is used much more intensively by users and is increasingly expected, [which] can only be fulfilled by tracking tools and marketing automation. [...] Of course, users also have a high interest in sufficient protection of their privacy. However, we are convinced that the data protection regulations in force in Germany [...] cover the expectations of most users.*"

Several respondents noted that they were already using external scanners to validate the security of their websites and that they commissioned professional security audits for their websites. Almost all respondents (A=7, B=5) forwarded the report to their IT departments for further analysis. There were also cases of respondents not being aware of alternatives to the privacy-invasive technologies they were using, with one respondent showing surprise when being informed about privacy-enhancing alternatives to their current practices.

**Complaints.** We received nine complaints, all of them from public insurers. As Group A contains a higher ratio of public insurers, we would have expected more complaints from this group. However, the majority of complaints came from Group B (A=3, B=6). Most complaints came from companies whose site had a poor rank in the initial scan (min 25, mean 80, median 87, max 123 of 125). Four of the nine complaints (A=1, B=3) were raised directly with the PrivacyScore team, i.e., not by replying to our solicitation mail. In one case, we received a positive reply to our solicitation email, while the PrivacyScore team received a complaint two days later. Eight of the nine companies requested to be excluded from future scans.

In the following, we will give an overview of the concerns raised in the complaints. We observed that the tone of the responses was generally more terse in Group B. Five of the six complaints from Group B explicitly referenced the ranking. One of the three complaints from Group A explicitly referenced the fact that the results were publicly available. One frequently raised point was that the scan was performed (B=2) and published (A=1, B=2) without asking for permission in advance. Three companies (A=1, B=2) claimed to be investigating if the scan constituted an illegal attack on their infrastructure, $B_{150}$ alleged a violation of competition law, and $A_{109}$ argued that the scan violated their copyright. In contrast, legal scholars concluded that PrivacyScore is fully compliant with German law [21].

Of particular interest is the complaint by $B_{150}$, which we will discuss in more detail here. After obtaining a low rank in the initial scan, their information security officer contacted the PrivacyScore team, asking to exclude the website from future scans. Almost two months later another response reached the PrivacyScore team, this time from the chief legal officer, stating that the legal department of the company had analyzed the case and had raised a number of issues with how the results were being displayed. In particular, they objected to a perceived incomprehensibility on how the rankings were computed and how old results from excluded websites were maintained. They claimed that this could be "*damaging to their company*", and may be illegal under the law against unfair competition. At the same time, they acknowledged that they were open to critical analysis of their website. In a subsequent phone call, the company explicitly mentioned the competitive and privacy-sensitive nature of the public health insurance market and the disadvantages a company could experience from a low rank in such a privacy ranking. A constructive discussion resulted in several changes to PrivacyScore being proposed and implemented, adding various clarifications to the results pages to make it easier for visitors to understand what they (do not) imply. In return, the insurance company started work to improve some of the security and privacy properties of its website. However, they still declined to be added back to the ranking.

**Other Responses.** We received 16 neutral reactions (A=8, B=8). Six organizations (A=4, B=2) explicitly stated that they were not interested in participating in the study. Five others (A=1, B=4) replied that our message had been forwarded internally, but we did not receive another response from them. Organization $A_{153}$ declined participation, citing insufficient capacity, and $B_{40}$ stated that responding would entail a work order to an external service provider. $A_{51}$ stated that they could not give any information on the topic of privacy. $B_{94}$ states that they did not understand the provided information. Finally, $A_{95}$ simply stated that they would not be making any changes to their website.

### 5.4 Changes to the Websites

To evaluate how websites changed over the course of the study, we performed measurements before sending our messages (T1) and repeated them at the end of the study (T2), skipping insurers that had asked to be excluded from future scans.

Twelve companies had stated that they were open to changing their website (A=6, B=6). However, at T2, only 4 (A=4) had actually made changes to one of the parameters measured by PrivacyScore. The total number of embedded third-party trackers (TPTs) increased over the course of the study period, from 411 to 439, which can be explained by a general inflation of the number of service providers in the market for behavioral advertising. We observe 12 (A=7, B=5) cases of at least one TPT being removed, and 13 (A=7, B=6) cases of at least one TPT being added.[7] Most of these websites belong to companies that did not respond to our messages. $A_{63}$, who had responded positively and promised to make changes, only replaced one TPT with a different one. $A_{91}$, who had provided us with a differentiated view of the trade-offs between privacy and economic success, appears to have briefly removed a number of third parties, but has since added them again. $A_{133}$ had stated their willingness to adapt their website – and indeed, they have removed all trackers, leaving only a cookie consent script hosted by a third party. Conversely, $A_{74}$ responded positively to our message, but only discussed the security aspects of the PrivacyScore evaluation, without regard to the privacy ratings. Their website added four additional trackers over the course of the study.

Several companies also made changes to the security of their systems by changing the configuration of their TLS setups. Four companies (A=3, B=1) disabled the outdated and insecure protocol version TLS 1.0, and two of them (A=1, B=1) also disabled TLS 1.1, leaving only the latest available version TLS 1.2 active. Five companies (A=3, B=2) enabled HTTP Strict Transport Security, while one (B=1) disabled it. However, none of them had responded to our message, so it is unknown if this happened as a reaction to our messages, or due to other, unrelated reasons. Similarly, 10 companies (A=6, B=4) started automatically forwarding all visitors to the secure version of their websites. Closer investigation revealed that while none of them had responded to our messages, five are maintained by the web design agency maintaining the website of $B_{128}$, who had asked to be excluded from future scans and threatened legal action. A manual visit of their website revealed that they, too, now forwarded all visitors to the secure version of their website. Thus, it is plausible that our messages were at least partially responsible for this agency-wide change.

Our observations indicate that insurers are willing to deploy changes that benefit the privacy interest of users without impacting the economic interests of the insurer (e.g., enabling a more recent TLS versions). If privacy and company interests are in conflict, companies are more reluctant to make a change (cf. the smaller number of insurers that removed third-party trackers from their site).

To put our results into context we compare the impact of our solicitation mails with the effect of regulatory changes (cf. Fig. 2). For this purpose, we scanned the websites once more in August 2018 (T3). Besides the introduction of the GDPR in May 2018, the timeframe also included new PCI DSS rules coming into effect in June 2018,[8] requiring websites that process credit card payments to update their TLS configuration.

Due to technical issues with PrivacyScore, scans for 10 sites reproducibly failed and had to be excluded in T3. Moreover, we observed significantly higher failure rates for the mail server TLS scans in T3 than in T1 and T2, precluding a meaningful comparison in this area. Cursory investigations indicate that the failures are due to the increasing prevalence of spam defense mechanisms deployed by mail server operators.

---

[7] We consider individual third parties, not the total number. Thus, if a website removes one third party and adds a different one, it will be counted in both groups.

[8] https://blog.pcisecuritystandards.org/are-you-ready-for-30-june-2018-sayin-goodbye-to-ssl-early-tls (accessed 2018-09-11)

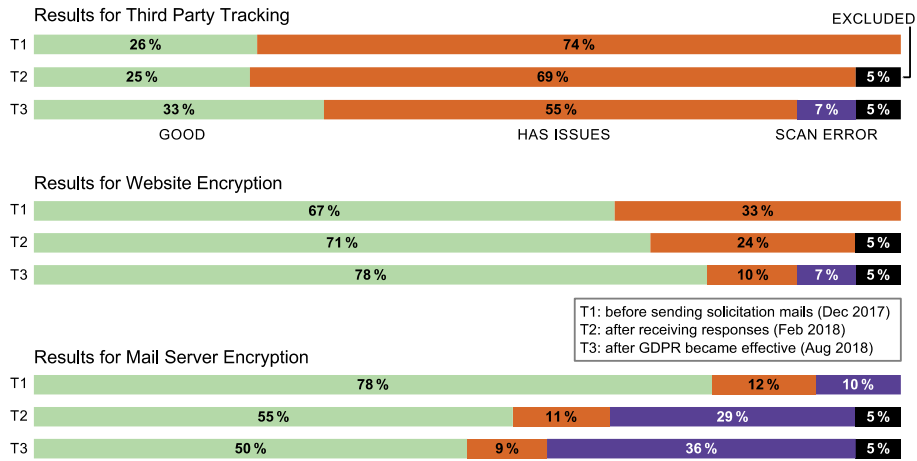

**Figure 2.** Scan results of health insurer websites at different points in time

As expected, the changes between T2 and T3 are substantial. 19 sites[9] removed all third parties, while 4 sites reintroduced trackers (compared to 2 and 1, respectively, in our study). 16 improved their TLS configuration (compared to 11 in our study), with 6 adding and 1 removing HSTS (5 additions and 1 removal in our study).[10]

Even though the time between T2 and T3 is much longer than between T1 and T2, it is unlikely that the observed differences are only due to the different durations given the changes to two regulatory frameworks. While we cannot infer what reasons served as an incentive for insurers to improve security and privacy features of their websites between T2 and T3, the combination of the passage of time and new regulatory requirements has a higher impact than our solicitation messages. This is to be expected, as the GDPR and PCI DSS changes were important and widely publicized, with strong incentives for compliance. While regulation has a large effect, transparency can still remain a valuable tool in affecting changes, as regulatory changes are infrequent events. Furthermore, regulation only establishes a lower bound of acceptable behavior, without incentives for exceeding the minimum requirements. Since the minimum requirements still permit many privacy-invasive techniques, transparency can serve as an incentive to exceed these requirements, which is vital for improving the state of online privacy.

## 6 Discussion

In this section, we discuss our results and their implications. The goal of our research was to investigate how health insurance companies react to transparency through public

---

[9] This includes 10 websites operated by the same association of insurers, whose websites are centrally managed. Thus, the number of distinct organizations making changes is at most 10.

[10] The difference with the changes visible in Figure 2 is due to some websites moving to the *excluded* or *failed* category

security and privacy ratings of their websites (RQ1) and if knowledge about the results being displayed in a ranking changes their response (RQ2).

We found that the responses varied greatly in the level of detail and content, ranging from detailed analyses of the trade-offs between user privacy and economic success to terse legal threats seeking removal from the public ranking. The responses cited a variety of reasons for the current state of their websites, ranging from conscious trade-offs between user privacy and the economic success of their company to technical limitations, e.g., third-party appliances whose configuration cannot be changed. One respondent also identified tension between the users' expectations of personalized content, which necessitates tracking, and user privacy.

### 6.1 Rankings Create Antagonism

We observed that public health insurers are more likely to complain than private insurance companies. Multiple complaints cited the privacy-sensitive nature of the health insurance market, and the potential competitive disadvantages caused by a bad privacy result. This led to a number of companies asking to be excluded from future scans, often under the threat of legal action based on questionable legal bases, including copyright, competition law, and cybercrime legislation [21].

We also observed a higher proportion of complaints from Group B (A=3, B=6), which was explicitly informed that their website was part of a ranking. This seems to indicate that the public ranking led to a higher probability of complaints being made. This theory is supported by the fact that five of the six complaints from Group B explicitly referenced the ranking in their complaints.

The complaints also have in common that all of them passed through the data protection and/or IT department of the companies, either as the initial point of contact for our study, or through company-internal forwarding. Thus, at first sight, a competing explanation for the higher ratio of complaints from Group B may be that the data protection and IT departments generally have a higher likelihood to complain. Coincidentally, we sent a higher fraction of our mails directly to data protection contacts in Group B (A=59%, B=68%), which corroborates the competing explanation. However, there are two arguments that challenge its validity: Firstly, several responders explicitly referenced the ranking in their complaints, and secondly, the complaint rate is still higher for Group B if only messages passing through data protection and/or IT departments are considered (A=21%, B=30%). Thus, we give more credence to our initial theory, which confirms RQ2.

### 6.2 Reasons for the Reluctance to Change

Our scans indicate that the websites under study are in permanent flux, with a general trend towards increasing the number of third-party trackers. As many of the insurers whose website changed during our study did not respond to our messages, the effect of our study on the websites is hard to quantify. Based on the received responses, we can attribute four changes directly to our messages, including one company that stopped using third-party trackers altogether, and assume that six more changes are at least partially the result of our messages. While these changes improve the privacy and security of website visitors, the low number of overall changes shows that the effectiveness of the messages was limited, indicating that at least in its current state, transparency and rankings through PrivacyScore do not significantly influence the willingness of most website operators to change their websites.

Many responses matched the *acquiescent approach* proposed in the IA [5], stating that they were in full compliance with applicable laws without giving additional details. A small number of companies moved towards the *proactive* approach by elaborating more on their internal processes, stating that tracking was disabled in critical areas, or citing frequent security audits of their websites.

We distil three reasons for the current state of insurance company websites: Conflicting value propositions, missing awareness, and negligence. Firstly, companies operate in a field of tension between their own economic goals (e.g., evaluating the effectiveness of their marketing campaigns), their reputation, and (sometimes conflicting) customer expectations like privacy and personalization. If not all expectations can be fulfilled, a trade-off needs to be found, leading companies to evaluate the costs of each solution. Stated in terms of the RbV [5], by gathering customer data (*intellectual resource*), companies can satisfy their own goals and some of the user expectations (e.g., personalization). As the use of web tracking is ubiquitous, it incurs almost no reputational cost and thus has low potential for differentiation (*relational resource*), leading most companies to pursue a *knowledge focus* and value their own economic goals higher than the privacy interests of their customers. This also explains their observed reluctance to be included in a ranking, making them favor an intentionally intransparent strategy, as proposed by Gerlach *et al.* [7].

Secondly, website operators may not be aware of alternative solutions that allow them to maintain the utility of their current solutions while decreasing their impact of the privacy of their users. Such alternative solutions include self-hosted tracking software like Matomo (matomo.org) that keeps the data under the control of the company, or two-click social media buttons (github.com/heiseonline/shariff) that do not disclose information to social networks on every page view.

Lastly, website operators may have been negligent and forgotten to configure their website correctly to respect the privacy of their users. This could manifest as the failure to enable the IP anonymization feature of Google Analytics (which is mandatory in Germany), or not forwarding visitors to the encrypted version of the website.

Problems caused by negligence can in many cases be remediated by a notification to the website operators, although such notifications have been shown to not always be reliable [17]. Raising awareness for privacy-preserving alternatives cannot easily be done at scale. In addition, awareness alone is not sufficient, as deciders need to be convinced that the benefits of switching to a privacy-preserving solution are worth the required effort and potential costs of changing the website, leading back to the issue of conflicting value propositions. These can only be influenced by changing the costs associated with the different options, which is easiest in the area of reputational costs. In the context of online privacy this means that privacy-invasive techniques have to become reputationally "expensive", which would turn forgoing their use into a relational resource, allowing differentiation through the *proactive approach*. This idea is at the center of the PrivacyScore project. However, our evaluation has shown that at the moment, the effect of transparency through PrivacyScore is not sufficient to cause large-scale changes. It is unclear if this is a general result or an artefact of the relatively unknown PrivacyScore platform. After all, companies may be more inclined to perform changes if a well-known consumer group or news outlet published a privacy ranking.

Finally, the behavior of companies can also be influenced by legislation and regulatory oversight. This was confirmed by our post-GDPR scan, which showed changes in a larger number of websites. Thus, the upcoming European ePrivacy regulation is a promising avenue for affecting further changes at scale.

## 7 Limitations, Future Research and Conclusion

Our study is subject to limitations. Firstly, we only investigate a single, privacy-sensitive sector – health insurance – in a single country. Extending and replicating the study with different sectors and in different countries could shed additional light on the general applicability of the results. Secondly, the number of respondents does not allow us to draw statistically significant conclusions. Our analysis also only considers two fixed dates when evaluating changes to the websites of the included companies, and we do not investigate if improvements are sustained over time or reverted.

Another limitation is the permeability of the group assignments. While we did not inform members of Group A about the ranking feature of PrivacyScore, the scan results page contains a (non-prominently presented) link to the ranking. Thus, members of Group A might have learned about it on their own. Nevertheless, we observed notable differences between Group A and B, which indicates that many members of Group A have not taken notice. The last limitation is the nature of PrivacyScore. As a relatively new platform, the publicity provided by it may present less of an incentive for change than a more popular and well-known platform or publication would have provided.

In conclusion, our results show that transparency – in the form of public assessments – can improve privacy features of websites. However, such efforts can also result in complaints and legal threats. A major factor limiting the willingness to change is the conflict between user privacy and the perceived need for privacy-invasive analytics for economic success: our solicitation mails led to much larger changes in areas where company and user interests are aligned, like website connection security. Another factor contributing to the current state appears to be a lack of awareness about privacy-preserving alternatives to common tracking services. While our study provided some initial insights on these difficulties, evaluating the effects of transparency on privacy remains a promising avenue for future work, for instance when publishing assessments in more widely disseminated channels like newspaper articles.

**Acknowledgements.** This work has been co-funded by the DFG as part of project C.1 within the RTG 2050 "Privacy and Trust for Mobile Users." The authors would like to thank Nora Wessels, Anne Laubach, Henning Pridöhl, and Pascal Wichmann for their assistance.